\documentstyle[aps,epsf,rotate]{revtex}
\tighten
\begin{document}
\title{ON GLOBAL STRUCTURE OF HADRONIC TOTAL CROSS SECTIONS}
\draft											  
\author{A.A.~Arkhipov}
\address{Institute for High Energy Physics\\
Protvino, 142284 Moscow Region, Russia}
\date{Received 17 December 1999}

\maketitle				   
\def\be{\begin{equation}}					 
\def\ee{\end{equation}}
\def\ber{\begin{eqnarray}}
\def\eer{\end{eqnarray}}	
\def\dint{\mathop{\intop\kern-0.5em\intop}}
\begin{abstract}
Simple theoretical formula describing the global structure of $pp$
and $p\bar p$ total cross sections in the whole range of energies
available up today has been derived. The fit to the experimental data
with the formula has been made. It is shown that there is a very good
correspondence of the theoretical formula to the existing
experimental data.  
\end{abstract}

 
\section{Introduction}
It is a well known fact that at energies above $\sqrt{s}\sim 20\,GeV$
all hadronic total cross sections rise with the growth of energy.
In 1970 the experiments at the Serpukhov accelerator revealed that
the $K^+p$ total cross section increased with energy \cite{1}.
Increase of the $pp$ total cross section has been discovered at the
CERN ISR \cite{2} and then the effect of rising total cross sections
was confirmed at the Fermilab accelerator \cite{3}.

Although nowadays we have in the framework of local quantum field
theory a gauge model of strong interactions formulated in terms of
the known QCD Lagrangian its relations to the so called ``soft"
(interactions at large distances) hadronic physics are far from
desired. Obviously the understanding of this physics is of high
interest because it has intrinsically fundamental nature. In spite of
more than 25 years after the formulation of QCD, we cannot still
obtain from the QCD Lagrangian the answer to the question why all
hadronic total cross sections grow with energy. We cannot predict
total cross sections in an absolute way starting from the fundamental
QCD Lagrangian as well mainly because it is not a perturbative
problem. We know e.g. that nonperturbative contributions to the gluon
propagator influence the behaviour of ``soft" hadronic processes and
the knowledge of the infrared behaviour of QCD is certainly needed to
describe the ``soft" hadronic physics in the framework of QCD.
Unfortunately, today we don't know the whole picture of the infrared
behaviour of QCD, we have some fragments of this picture though (see
e.g. Ref. \cite{4}).

At the same time it is more or less clear now that the rise of the
total cross sections is just the shadow (not antishadow!) of particle
production. 

Through the optical theorem the total cross section is related to the
imaginary part of the elastic scattering amplitude in the forward
direction. That is why the theoretical understanding of elastic
scattering is of the fundamental importance. From the unitarity
relation it follows also that the imaginary part of the elastic
scattering amplitude contains the contribution of all possible
inelastic channels in two-particle interaction. It is clear therefore
that we cannot understand elastic scattering without understanding
inelastic interaction.

A variety of different approaches to the dynamics of ``soft" hadronic
processes can conditionally be divided into two groups corresponding
to different forms of strong interaction dynamics: t-channel form and
s-channel one.

In the framework of t-channel form of the dynamics the popular Regge
phenomenology represents elastic and inelastic diffractive scattering
by the exchange of the Pomeron, a color singlet Reggeon with quantum
numbers of the vacuum. It should be noted, that the definition of the
Pomeron as Reggeon with the highest Regge trajectory $\alpha_P(t)$,
that carries the quantum numbers of the vacuum, is not only
one.\footnote{For the supercritical Pomeron $\alpha_P(0) - 1 = \Delta
\ll 1,\,  \Delta > 0$ is responsible for the growth of hadronic cross
sections with energy.} There are many other definitions  of the
Pomeron: Pomeron is a gluon ``ladder"  \cite{5}; Pomeron is a bound
state of two reggezied gluons -- BFKL-Pomeron \cite{6}; soft and hard
Pomerons \cite{7,8}, etc. This leapfrog is because of the exact
nature of the Pomeron and its detailed substructure remains such as
that anyone doesn't know what it is. The difficulty of establishing
the true nature of the Pomeron in QCD is almost obviously related to
the calculations of non-perturbative gluon exchange.

Nevertheless in the near past simple formulae of the Regge
phenomenology provided good parameterization of experimental data on
``soft" hadronic physics and pragmatic application of Pomeron
phenomenology had been remarkably successful (see e.g. last issue of
the Review of Particles Properties).

That was the case before the appearance of CDF data on single
diffractive dissociation \cite{9,10} and recent results from HERA
\cite{11}, which had shown that a popular model of supercritical
Pomeron did not describe new experimental data. Obviously, the
foundations of the Pomeron model require further theoretical study
and the construction of newer, more general phenomenological
framework, which would enable one to remove the discrepancy between
the model predictions and the experiment. Of course, it is good that
we have simple and compact form for representing a great variety of
data for different hadronic processes, but it is certainly bad that
power behaved total cross sections violate unitarity. Often and often
encountered claim, that the model with power behaved total cross
sections is valid in the non-asymptotic domain which has been
explored up today, is not correct because supercritical Pomeron model
is an asymptotic one by definition.

We suggested a new approach to the dynamical description of
one-particle inclusive reactions \cite{12}. The main point of our
approach is that new fundamental three-body forces are responsible
for the dynamics of particle production processes of inclusive type.
Our consideration revealed several fundamental properties of
one-particle inclusive cross-sections in the region of diffraction
dissociation. In particular, it was shown that the slope of the
diffraction cone in $p\bar p$ single diffraction dissociation was
related to the effective radius of three-nucleon forces in the same
way as the slope of the diffraction cone in elastic $p\bar p$
scattering was related to the effective radius of two-nucleon
forces. It was also demonstrated that the effective radii of two- and
three-nucleon forces, which were the characteristics of elastic and
inelastic interactions of two nucleons, defined the structure of the
total cross-sections in a simple and physically clear form. We
made an attempt to check up the structure on its correspondence to
the existing experimental data on proton-proton and proton-antiproton
total cross sections. It is a remarkable fact, which is presented in
this paper, that there is a very good correspondence.

First of all it should be elucidated what the three-body forces are.
It will be made in the next sections.

\section{Three-body forces in relativistic quantum theory}

Using the LSZ or the Bogoljubov reduction formulae in quantum field
theory \cite{13}, we can easily obtain the following cluster
structure for $3\rightarrow 3$ scattering amplitude (see Fig. 1)
\be
{\cal F}_{123} = {\cal F}_{12} + {\cal F}_{23} + {\cal F}_{13} + 
{\cal F}_{123}^C \label{2}
\ee
where ${\cal F}_{ij} , (i,j = 1,2,3)$ are $2 \rightarrow 2$
scattering  amplitudes, ${\cal F}_{123}^C$ is called the connected
part of the $3 \rightarrow 3$ scattering amplitude.

\vskip0.4cm
\begin{center}
\begin{picture}(370,30)
\thicklines
\put(0,0){\line(1,0){60}}
\put(0,30){\line(1,0){60}}
\put(30,15){\circle{30}}
\put(0,15){\line(1,0){15}}
\put(45,15){\line(1,0){15}}

\put(65,12){\mbox{=}}

\put(75,0){\line(1,0){60}}
\put(75,15){\line(1,0){60}}
\put(105,22,5){\circle{15}}
\put(75,30){\line(1,0){60}}

\put(140,12){\mbox{+}}

\put(150,0){\line(1,0){60}}
\put(150,15){\line(1,0){60}}
\put(150,30){\line(1,0){60}}
\put(180,7.5){\circle{15}}

\put(215,12){\mbox{+}}

\put(230,0){\line(1,0){60}}
\put(230,30){\line(1,0){60}}
\put(260,15){\circle{30}}
\put(260,15){\oval(40,20)[b]}
\put(230,15){\line(1,0){10}}
\put(280,15){\line(1,0){10}}

\put(300,12){\mbox{+}}

\put(315,0){\line(1,0){60}}
\put(315,30){\line(1,0){60}}
\put(345,15){\circle{30}}
\put(315,15){\line(1,0){15}}
\put(360,15){\line(1,0){15}}

\put(335,7){\mbox{\huge C}}

\end{picture}

\bigskip
\centerline{Figure 1: Cluster structure for $3\rightarrow 3$
scattering amplitude.}
\end{center}

\bigskip
In the framework of single-time formalism in quantum field theory
\cite{14} we construct the $3 \rightarrow 3$ off energy shell
scattering amplitude $T_{123}(E)$ with the same (cluster) structure 
as (\ref{2})
\be
T_{123}(E) = T_{12}(E) + T_{23}(E) + T_{13}(E) + T_{123}^C(E).
\label{3}
\ee
Following the tradition we'll call the kernel describing the
interaction of three particles as the three particle interaction
quasipotential. The three particle interaction quasipotential
$V_{123}(E)$ is related to the off-shell $3 \rightarrow 3$ scattering
amplitude $T_{123}(E)$ by the Lippmann-Schwinger type equation
\be
T_{123}(E) = V_{123}(E) + V_{123}(E)G_0(E)T_{123}(E). \label{4}
\ee
There exists the same transformation between two particle
interaction quasi\-potentials $V_{ij}$ and off energy shell $2
\rightarrow 2$ scattering amplitudes $T_{ij}$
\be
T_{ij}(E) = V_{ij}(E) + V_{ij}(E)G_0(E)T_{ij}(E). \label{5}
\ee
It can be shown that in the quantum field theory the three particle
interaction quasipotential has the following structure
\cite{15}:
\be
V_{123}(E) = V_{12}(E) + V_{23}(E) + V_{13}(E) + V_0(E). \label{6}
\ee
The quantity $V_0(E)$ is called the three-body forces quasipotential.
The $V_0(E)$ represents the defect of three particle interaction
quasipotential over the sum of two particle interaction
quasipotentials and describes the true three-body interactions.
Three-body forces quasipotential is an inherent connected part of
total three particle interaction quasipotential which can not be
represented by the sum of pair interaction quasipotentials.

The three-body forces scattering amplitude is related to the
three-body forces quasipotential by the equation
\be
T_0(E) = V_0(E) + V_0(E)G_0(E)T_0(E). \label{7}
\ee

It should be stressed that the three-body forces appear as a 
result of consistent consideration of three-body problem in the
framework of local quantum field theory.

\section{Global analyticity of the three-body forces} 

Let us introduce the following useful notations:
\be
<p'_1 p'_2 p'_3\vert S - 1\vert p_1 p_2 p_3> =
2\pi i\delta ^4(\sum_{i=1}^{3}p'_i-\sum_{j=1}^{3}p_j)
{\cal F}_{123}(s;{\hat e}',\hat e), \label{8}
\ee
\[
s = (\sum_{i=1}^{3}p'_i)^2 = (\sum_{j=1}^{3}p_j)^2.
\]
${\hat e}', \hat e \in S_5$ are two unit vectors on five-dimensional
sphere describing the configuration of three-body system in initial
and final states (before and after scattering).

We will denote the quantity $T_0$ restricted on the energy shell as
\[
T_0\mid_{on\, energy\, shell}\, = {\cal F}_0.
\]
The unitarity condition for the quantity ${\cal F}_0$ with 
account for the introduced notations can be written in form
\cite{16,17}
\be
Im{\cal F}_0(s;{\hat e}',\hat e) = \pi A_3(s)\int d\Omega _{5}({\hat
e}''){\cal F}_0(s;{\hat e}',{\hat e}'')
\stackrel{*}{\cal F}_0(s;\hat e,{\hat e}'') + H_0(s;{\hat e}',\hat
e),
\label{9}
\ee

\[
Im{\cal F}_0(s;{\hat e}',\hat e)=\frac{1}{2i}\left[{\cal F}_0(s;{\hat
e}',\hat e)
-\stackrel{*}{\cal F}_0(s;\hat e,{\hat e}')\right],
\]
where
\[
A_3(s) = {\Gamma}_3(s)/S_5 ,
\]
${\Gamma}_3(s)$ is the three-body phase-space volume, $S_5$ is the
volume of unit five-dimensional sphere. $H_0$ defines the
contribution of all inelastic channels emerging due to three-body
forces.

Let us introduce a special notation for the scalar product of two
unit vectors 
${\hat e}'$ and $\hat e$
\be
\cos \omega = {\hat e}^{'}\cdot \hat e. \label{10}
\ee
We will use the other notation for the three-body forces scattering
amplitude as well 
\[
{\cal F}_0(s;{\hat e}',\hat e) = {\cal F}_0(s;\eta,\cos\omega),
\]
where all other variables are denoted through $\eta$.

Now we are able to go to the formulation of our basic assumption
on the analytical properties for the three-body forces scattering
amplitude 
\cite{16,17}.

We will assume that for physical values of the variable $s$ and fixed 
values of  $\eta$ the amplitude ${\cal F}_0(s;\eta,\cos\omega)$ is an 
analytical function of the variable $\cos\omega$ in the ellipse
$E_0(s)$ 
with the semi-major axis
\be
z_0(s) = 1 + \frac{M_0^2}{2s} \label{11}
\ee
and for any $\cos\omega \in E_0(s)$ and physical values of $\eta$ it
is polynomially bounded in the variable $s$. $M_0$ is some constant
having mass dimensionality.

Such analyticity of the three-body forces amplitude was called a
global one. The global analyticity may be considered as a direct
geometric generalization of the known analytical properties of
two-body scattering amplitude strictly proved in the local quantum
field theory \cite{18,19,20,21,22}.

At the same time the global analyticity results in the generalized
asymptotic bounds. For example, the generalized asymptotic bound for
$O(6)$-invariant three-body forces scattering amplitude looks like
\cite{16,17}
\begin{equation}
Im\,{\cal F}_0(s;...) \leq \mbox{Const}\, s^{3/2} 
\bigl(\frac{\ln s/s'_0}{M_0}\bigr)^5 = \mbox{Const}\,
s^{3/2}R_0^5(s),
\label{12}
\end{equation}
where $R_0(s)$ is the effective radius of the three-body forces
introduced according to \cite{22}, where the effective radius
of two-body forces has been defined
\be
R_0(s) = \frac{\Lambda_0}{\Pi(s)} = \frac{r_0}{M_0}\ln
\frac{s}{s'_0},
\quad \Pi(s) = \frac{\sqrt{s}}{2},\quad s\rightarrow \infty,
\label{13}
\ee
$r_0$ is defined by the power of the amplitude ${\cal F}_0$ growth 
at high energies \cite{17}, $M_0$ defines the semi-major axis of the
global analyticity ellipse (\ref{11}), ${\Lambda}_0$ is the effective
global orbital momentum, $\Pi(s)$ is the global momentum of
three-body system, $s'_0$ is a scale defining unitarity saturation of
three-body forces.
 
It is well known that the Froissart asymptotic bound \cite{23} can
be experimentally verified because with the help of the optical
theorem we can  connect the imaginary part of $2 \rightarrow 2$
scattering amplitude with the experimentally measurable quantity
which is the total cross section. So, if we want to have a
possibility for the experimental verification of the generalized
asymptotic bounds $(n\geq3)$ we have to establish a connection
between the many-body forces scattering  amplitudes and the
experimentally measurable quantities. For this aim we have considered
the problem of high energy particle scattering from deuteron and  on
this way we found the connection of the three-body forces scattering
amplitude with the experimentally measurable quantity which is the
total cross section for scattering from deuteron \cite{24}. Moreover, 
the relation of the three-body forces scattering amplitude to
one-particle inclusive cross sections has been established \cite{25}. 

We shall briefly sketch now the basic results of our analysis of high
energy particle scattering from deuteron. 

\section{Scattering from deuteron}
The problem of scattering from two-body bound states was treated in
works \cite{24,25} with the help of dynamic equations obtained on the
basis of single-time formalism in QFT \cite{15}. As has been shown in
\cite{24,25}, the total cross section in scattering from deuteron
can  be expressed by the formula
\be
\sigma_{hd}^{tot}(s) = \sigma_{hp}^{tot}(\hat s)
+\sigma_{hn}^{tot}(\hat s) -
\delta\sigma(s), \label{14}
\ee
where $\sigma_{hd}, \sigma_{hp}, \sigma_{hn}$  are the total cross
sections
in scattering from deuteron, proton and neutron, 
\begin{equation}
\delta\sigma(s) = \delta\sigma_G(s) +\delta\sigma_0(s),\label{15}
\end{equation}
\begin{equation}
\delta\sigma_G(s) =
\frac{\sigma_{hp}^{tot}(\hat s) \sigma_{hn}^{tot}(\hat s)}{4\pi(
R^2_d+B_{hp}(\hat s)+B_{hn}(\hat s))} 
\equiv \frac{\sigma_{hp}^{tot}(\hat s) \sigma_{hn}^{tot}(\hat
s)}{4\pi
R^2_{eff}(s)},\ \ \hat s = \frac{s}{2},
\label{16}
\end{equation}
$B_{hN}(s)$ is the slope of the forward diffraction peak in the
elastic scattering from nucleon, $1/R_d^2$ is defined by the deuteron
relativistic formfactor
\be
\frac{1}{R_d^2} \equiv \frac{q}{\pi }\int\frac{d\vec \Delta \Phi
(\vec \Delta )}{2\omega_h (\vec 
q+\vec \Delta )}\delta \left[\omega_h (\vec q+\vec \Delta )-\omega_h
(\vec q\,)\right],\ \ \frac{s}{2M_d}\cong q\cong \frac{\hat
s}{2M_N}, \label{17}
\ee
$\delta\sigma_G$ is the Glauber correction or shadow effect. The
Glauber shadow correction originates from elastic rescatterings of an
incident particle on the nucleons inside the deuteron.

The quantity $\delta\sigma_0$ represents the contribution of the
three-body forces to the total cross section in scattering from
deuteron. The physical reason for the appearance of this quantity is
directly connected with the inelastic  interactions of an incident
particle with the nucleons of deuteron. Paper \cite{25} 
provides for this quantity the following expression:
\begin{equation}
\delta\sigma_0(s) = -\frac{(2\pi)^3}{q}
\int \frac{d\vec\Delta\Phi(\vec\Delta)}
{2E_p(\vec\Delta /2)2E_n(\vec\Delta /2)}
Im\,R\bigl(s;-\frac{\vec\Delta}{2},\frac{\vec\Delta}{2},\vec q; 
\frac{\vec\Delta}{2},-\frac{\vec\Delta}{2},\vec q \bigr),\label{18}
\end{equation}
where $q$ is the incident particle momentum in the lab system (rest
frame of deuteron), $\Phi(\vec\Delta)$ is the deuteron relativistic
formfactor, normalized to unity at zero, 
\[
E_N(\vec\Delta)=\sqrt{\vec\Delta^2 + M^2_N}\quad N = p, n,
\]
$M_N$ is the nucleon mass. The function $R$ is expressed via the
amplitude of the three-body forces $T_0$ and the amplitudes of
elastic scattering from the nucleons $T_{hN}$ by the relation
\begin{equation}
R = T_0 + \sum_{N=p,n}(T_0G_0T_{hN} + T_{hN}G_0T_0).\label{19}
\end{equation}
In \cite{24} the contribution of three-body forces to the
scattering amplitude from deuteron was related to the processes of
multiparticle production in the inelastic interactions of the
incident particle with the nucleons of deuteron. This was done with
the help of the unitarity  equation. The character of the energy
dependence of $\delta\sigma_0$ was shown to be governed by the energy
behaviour of the corresponding inclusive cross sections.

Here for simplicity let us consider the model where the imaginary
part of the three-body forces scattering amplitude has the form
\begin{equation}
Im\,{\cal F}_0(s; \vec p_1, \vec p_2, \vec p_3; \vec q_1, \vec q_2,
\vec q_3)
= f_0(s)
\exp \Biggl\{-\frac{R^2_0(s)}{4} \sum^{3}_{i=1} (\vec p_i-\vec
q_i)^2\Biggr\},\label{20}
\end{equation}
where $f_0(s)$, $R_0(s)$ are free parameters which in general may
depend on the total energy of three-body interaction. Note that the
quantity $f_0(s)$ has the dimensionality $[R^2]$. The model
assumption (\ref{20}) is not significant for our main conclusions but
allows one to make some calculations exactly in a closed form.

In case of unitarity saturation of the three-body forces, we have 
from the generalized asymptotic theorems 
\be
f_0(s) \sim \mbox{Const}\, s^{3/2}{\Bigl(\frac{\ln
s/s'_0}{M_0}\Bigr)}^5 = \mbox{Const}\, s^{3/2}R_0^5(s),\label{21}
\ee
\be
R_0(s) = \frac{r_0}{M_0} \ln s/s'_0 \quad s\rightarrow
\infty.\label{22}
\ee

In the model all the integrals can be calculated in the
analytical form. As a result we obtain for the quantity
$\delta\sigma_0$ \cite{25}
\be
\delta\sigma_0(s) = \frac{(2\pi)^{6}f_0(s)}{sM_N }
\Biggl\{\frac{\sigma_{hN}(s/2)}{2\pi[B_{hN}(s/2)+R^2_0(s)-R^4_0(s)/4(
R^2_
0(s)+R^2_d)]}-1\Biggr\}
\frac{1}{[2\pi(R^2_d+R^2_0(s))]^{3/2}}.\label{23}
\ee 
If the condition
\be
R_0^2(s) \simeq B_{hN}(s/2) \ll R_d^2 \label{24}
\ee
is realized, then we obtain from expression (\ref{23})
\be
\delta\sigma_0(s) =
(2\pi)^{9/2}\frac{f_0(s)\chi(s)}{sM_NR_d^3},\label{25}
\ee
where
\be
\chi(s) = \frac{\sigma_{hN}^{tot}(s/2)}{2\pi[B_{hN}(s/2)+R_0^2(s)]} -
1,\label{26}
\ee
and we suppose that asymptotically
\[
B_{hp}=B_{hn}\equiv B_{hN},\quad \sigma_{hp}^{tot}=\sigma_{hn}^{tot}
\equiv \sigma_{hN}^{tot}.
\]
It should be noted that (\ref{24}) is justified by the physical
conditions which are realized in the problem of high-energy particle
scattering from deuteron at the recently available energies. Of
course, there is some energy where (\ref{24}) can no longer be true
and this situation can be considered as well. We can calculate this
energy. From an experiment it is known that $R_d^2 \cong 100
GeV^{-2}$. Then we can easily calculate (see Eqs. (47,48) later) that
$2B_{hN}(s_1)=R_d^2$ at $\sqrt{s_1}=1515.92\, TeV$. In the same way
(see Eq. (50) later) we can find that $R_0^2(s_2)=R_d^2$ at
$\sqrt{s_2}=77.43\, PeV$. So, both $s_1$ and $s_2$ are extremely high
energies.

In the 50th Bogoljubov proposed the idea to describe stable compound
systems by local fields. Bogoljubov's idea has brilliantly been
realized by Zimmermann in his famous paper \cite{26}. Zimmermann's
construction for local deuteron field looks like
\be
B_d(X) = \lim_{x \to 0}\,\frac{ T\left(\Phi _p(X+\frac{1}{2}x)\Phi
_n(X-\frac{1}{2}x)
\right)}{<0\vert T\left(\Phi _p(\frac{1}{2}x)\Phi _n(-\frac{1}{2}x)
\right) \vert d >}.\label{27}
\ee
It has been proved \cite{26} that $B_d(X)$ satisfies microlocal
causality. Moreover, the asymptotic deuteron fields constructed with 
the Yang--Feldman  procedure fulfil the commutation relations of
Fock representation.  The hadron--deuteron scattering amplitude can
be presented by LSZ reduction formula in the form
\[
< {\vec P}_d{\vec p}_h\vert S-1\vert  {\vec Q}_d{\vec q}_h> = 
 \dint dXdx\dint dYdy
{\stackrel{*}{f}}_{M_d;{\vec P}_d}(X)
{\stackrel{*}{f}}_{m_h;{\vec p}_h}(x)\times 
\]
\be
{\stackrel{\rightarrow}{K}}_{X}^{M_d}
{\stackrel{\rightarrow}{K}}_{x}^{m_h}
<0\vert T\left(B_d(X)\Phi _h(x)B_d(Y)\Phi _h(y)\right)\vert 0>
{\stackrel{\leftarrow}{K}}_{Y}^{M_d}
{\stackrel{\leftarrow}{K}}_{y}^{m_h}
f_{M_d;{\vec Q}_d}(Y)f_{m_h;{\vec q}_h}(y) \label{28}
\ee
where $K_x^m \equiv  {\framebox(8,8){}}\,_x+m^2$ is Klein-Gordon-Fock 
differential operator,
\[
f_{m;\vec p}\,(x) = (2\pi )^{-3/2}exp(-ipx),\quad
p^0 = E(\vec p,m) = \sqrt{{\vec p}\,^2+m^2}.
\]
Of course, the construction of local interpolating Heisenberg fields
is not a unique procedure. There are equivalence classes of different
fields (Borchers's classes \cite{27}), which have the same asymptotic
fields and give rise to the same S-matrix. Anyway Zimmermann's
construction allows us to use the local quantum field theory
Causality-Spectrality--Analyticity-Unitarity (CS--AU) machine
\cite{13} and prove the Froissart theorem for hadron--deuteron
elastic scattering amplitude as well and, as a consequence, obtain
the Froissart bound for total cross section in hadron--deuteron
interaction. 

First two terms in Eq. (\ref{14}) fulfil the Froissart bound. The
quantity $\delta\sigma_G$ meets a stronger bound than the
Froissart one because we know that 
\[
\sigma_{hN}^{tot}(s,s_0) \sim \ln^2(s/s_0)\Longrightarrow
B_{hN}(s,s_0)\sim \ln^2(s/s_0),
\]
and, therefore, we have from Eq. (\ref{16})\footnote{The bound
$\delta\sigma_G(s) < 2\sigma_{hN}^{el}(\hat s)$ is also true when
$\rho_{el}\not=0.$}
\[
\delta\sigma_G(s) < 2\sigma_{hN}^{el}(\hat s), \quad
(\sigma_{hN}^{el}(s) =
\frac{\sigma_{hN}^{tot}(s)^2}{16\pi B_{hN}(s)}, \ \ \rho_{el}=0).
\]
Obviously, from Eq. (\ref{16}) it follows that the bound
$\delta\sigma_G(s)<2\sigma_{hN}^{el}(\hat s)$ is true at any relation
between $B_{hN}$ and $R_d^2$. But if the condition (\ref{24}) is
realized we have a more stronger bound
\[
\delta\sigma_G(s)<2\epsilon(\hat s)\sigma_{hN}^{el}(\hat s), \quad
\epsilon(\hat s)=2B_{hN}(\hat s)/R_d^2\ll 1.
\]
From expression (\ref{25}) for the correction $\delta\sigma_0$ in the
case of unitarity saturation of the three-body forces with the energy
dependence given by Eq. (20) then it follows that
\[
\delta\sigma_0(s) < C_0\,ln^2s \quad (C_0 < C_{hN}^F=\pi/m_{\pi}^2)
\quad s\rightarrow\infty 
\]
if and only if the asymptotic bound
\begin{equation} 
\chi(s) <  \frac{C}{\sqrt{s}{\ln}^3 s},\quad s\rightarrow 
\infty. \label{29}
\end{equation} 
is valid. 

\section{Global structure of hadronic total cross sections}

Let's rewrite the equation for $\chi(s)$ 
\[
\chi(s) = \frac{\sigma_{hN}^{tot}(s/2)}{2\pi[B_{hN}(s/2)+R_0^2(s)]} -
1
\]
in the form

\be
\sigma_{hN}^{tot}(s) = 2\pi\left[B_{hN}(s) +
R_0^2(2s)\right]\left(1+\chi\right).\label{30}
\ee
From the Froissart and generalized asymptotic bounds we have
\[
\chi(s) = O\left(\frac{1}{\sqrt{s}\ln^3s}\right),\quad s\rightarrow
\infty.
\]
We also know that \cite{20}
\be
\sigma_{hN}^{tot}(s,s_0) \sim \ln^2(s/s_0)\Longrightarrow
B_{hN}(s,s_0)\sim \ln^2(s/s_0),\label{31}
\ee
and Eq. (\ref{30}) gives
\[
R_0^2(2s,s'_0)\sim \ln^2(2s/s'_0)\sim \ln^2(s/s_0),\quad
s\rightarrow\infty.
\]
Therefore we come to the following asymptotic consistency condition
\be
\fbox{$\displaystyle s'_0 = 2s_0$}\,.\label{32}
\ee
The asymptotic consistency condition tells us that we have not any
new scale. The scale defining unitarity saturation of three-body
forces is unambiguously expressed by the scale which defines
unitarity saturation of two-body forces. In that case we have
\[
R_0^2(2s,s'_0) = R_0^2(s,s_0)
\]
and
\be
\fbox{$\displaystyle \sigma_{hN}^{tot}(s) = 2\pi\left[B_{hN}(s) +
R_0^2(s)\right]\left(1+\chi(s)\right)$} \label{33}
\ee
with a common scale $s_0$. 

Reminding the relation between the effective radius of two-body
forces and the slope of diffraction cone in elastic scattering 
\be
B_{hN}(s) = \frac{1}{2}R_{hN}^2(s) \label{34}
\ee
we obtain
\be
\sigma_{hN}^{tot}(s) = \pi R_{hN}^2(s) +
2\pi R_0^2(s),\quad s\rightarrow\infty.\label{35}
\ee
Equations (\ref{33}) and (\ref{35}) define new nontrivial structure
of hadronic total cross section. It should be emphasized that the
coefficients staying in the R.H.S. of Eq. (\ref{35}) in front of
effective radii of two- and three-body forces are strongly fixed.

It is useful to compare the new structure of total hadronic cross
section with the known one. We have from unitarity

\be
\sigma_{hN}^{tot}(s) = \sigma_{hN}^{el}(s) +
\sigma_{hN}^{inel}(s).\label{36}
\ee
If we put
\be
\sigma_{hN}^{el}(s) = \pi R_{hN}^{el\,^2}(s),\quad
\sigma_{hN}^{inel}(s) = 2\pi R_{hN}^{inel\,^2}(s),\label{37}
\ee
then we come to the similar formula

\be
\sigma_{hN}^{tot}(s) = \pi R_{hN}^{el\,^2}(s) + 2\pi
R_{hN}^{inel\,^2}(s).\label{38}
\ee
But it should be borne in mind
\be
R_{hN}^2(s) \not= R_{hN}^{el\,^2}(s),\quad R_0^2(s) \not=
R_{hN}^{inel\,^2}(s).\label{39}
\ee
In fact we have
\be
\sigma_{hN}^{el}(s) = \frac{\sigma_{hN}^{tot\,^2}(s)}{16\pi
B_{hN}(s)}
= \frac{\sigma_{hN}^{tot\,^2}(s)}{8\pi R_{hN}^2(s)},\label{40}
\ee
\be
\sigma_{hN}^{inel}(s) = \sigma_{hN}^{tot}(s)\left[1 -
\frac{\sigma_{hN}^{tot}(s)}{8\pi R_{hN}^2(s)}\right].\label{41}
\ee
Of course, Eqs. (\ref{37}) are the definitions of $R_{hN}^{el}$ and
$R_{hN}^{inel}$. The definition of $R_{hN}^{el}$ corresponds to our
classical imagination, the definition of $R_{hN}^{inel}$
corresponds to our knowledge of quantum mechanical problem for
scattering from a black disk. Let us suppose that
\be
\sigma_{hN}^{tot}(s_m) \cong  \pi R_{hN}^2(s_m),\ \
s_m\in{\cal M},\quad \left(R_0^2(s_m)\ll
R_{hN}^2(s_m)\right),\label{42}
\ee
then we obtain
\be
\sigma_{hN}^{el}(s_m) = \frac{1}{8}\pi R_{hN}^2(s_m),\quad
\sigma_{hN}^{inel}(s_m) = \frac{7}{8}\pi R_{hN}^2(s_m).\label{43}
\ee
This simple example shows that the new structure of total hadronic
cross sections is quite different from that given by Eq.
(\ref{36}). The reason is that structure (\ref{33}) is of dynamical
origin. We have mentioned above that the coefficients, staying in
R.H.S. of Eq. (\ref{35}) in front of effective radii of two- and
three-body forces, are strongly fixed. In fact, we found here the 
answer to the old question: Why the constant ($\pi/m_{\pi}^2 \approx
60\,mb $) staying in the Froissart bound is too large in the light of
existing experimental data. The constant in R.H.S. of Eq. (\ref{35}),
staying in front of the effective radius of hadron-hadron
interaction, is 4 times smaller than the constant in the Froissart
bound. But this is too small to correspond to the experimental data.
The second term in R.H.S. of Eq. (\ref{35}) fills an emerged gap.
Besides from the Froissart bound
\[
\sigma_{hN}^{tot}(s) \leq 4\pi R_{hN}^2(s)
\]
we obtain the bound for the effective radius of three-body forces
\be
R_0^2(s) \leq \frac{3}{2}R_{hN}^2(s). \label{44}
\ee
It is a remarkable fact that the quantity $R_0^2$, which has the
clear physical interpretation, at the same time, is related to the
experimentally measurable quantity which the total cross section is.
This important circumstance gives rise to the nontrivial
consequences \cite{28}.

We made an attempt to check up the structure (\ref{33}) on its
correspondence to the existing experimental data. Our results are
presented below.

At the first step, we made a weighted fit to the experimental data on
the proton-antiproton total cross sections in the range 
$\sqrt{s}>10\, GeV$. The data were fitted with the function of the
form predicted by the Froissart bound in the spirit of our
approach\footnote{Recently, from a careful analysis of the
experimental data and comparative study of the known characteristic
parameterizations, Bueno and Velasco have shown that statistically a
``Froissart-like" type parameterization for proton-proton and
proton-antiproton total cross sections is strongly favoured
\cite{29}.}
\be
\sigma^{tot}_{asmpt} = a_0 + a_2 \ln^2(\sqrt{s}/\sqrt{s_0})
\label{45}
\ee
where $a_0, a_2, \sqrt{s_0}$ are free parameters. We accounted for
the experimental errors $\delta x_i$ (statistical and systematic
errors added in quadrature) by fitting to the experimental points
with the weight $w_i=1/(\delta x_i)^2$. Our fit yielded
\be
a_0 = (42.0479\pm 0.1086)mb,\quad a_2 = (1.7548\pm
0.0828)mb,\label{46}
\ee
\be
\sqrt{s_0} = (20.74\pm 1.21)GeV.\label{47}
\ee
The fit result is shown in Fig. 2.
\begin{center}
\begin{picture}(288,198)
\put(15,10){\epsfbox{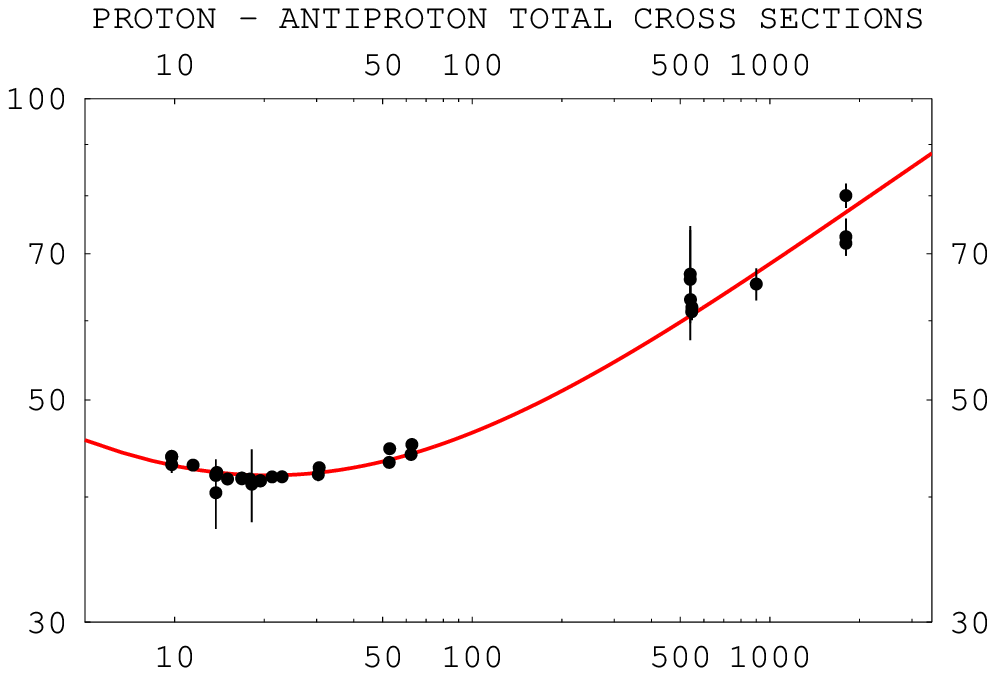}}
\put(144,0){$\sqrt{s}\, (GeV)$}
\put(0,77){\rotate{\large$\sigma_{tot} (mb)$}}
\end{picture}
\end{center}
Figure 2: The total proton-antiproton cross sections versus
$\sqrt{s}$
compared with formula (\ref{45}). Solid line represents our fit
to the data. Statistical and systematic errors added in quadrature.

\bigskip
After that we made a weighted fit to the experimental data on the
slope of diffraction cone in elastic $p\bar p$ scattering. The
experimental points and the references, where they have been
extracted from, are listed in \cite{30}. The fitted function of
the form
\be
B = b_0 + b_2 \ln^2(\sqrt{s}/20.74), \label{48}
\ee
which is also suggested by the asymptotic theorems of local quantum
field theory, has been used. The value $\sqrt{s_0}$ was fixed by
(\ref{47}) from the fit to the $p\bar p$ total cross sections data.
Our fit yielded
\be
b_0 = (11.92\pm 0.15)GeV^{-2} ,\quad b_2 = (0.3036\pm
0.0185)GeV^{-2}.\label{49} 
\ee
The fitting curve is shown in Fig. 3.

\bigskip
\begin{center}
\begin{picture}(288,188)
\put(15,10){\epsfbox{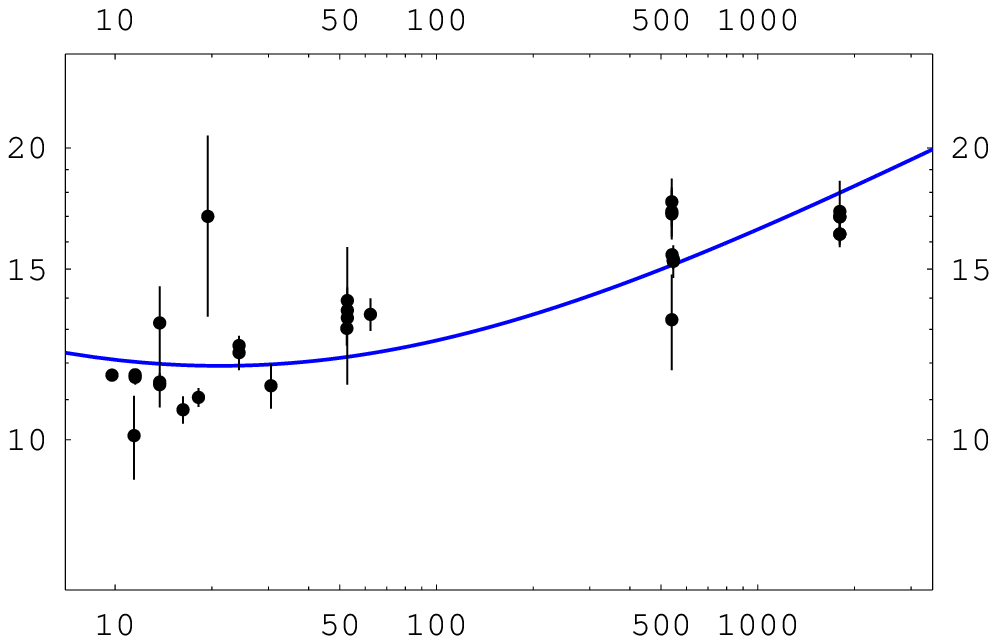}}
\put(144,0){$\sqrt{s}\, (GeV)$}
\put(0,77){\rotate{$B\ \ (GeV^{-2})$}}
\end{picture}
\end{center}
Figure 3: Slope $B$ of diffraction cone in $p\bar p$
elastic scattering. Solid line represents our fit to the data.

\bigskip
At the final stage we build a global (weighted) fit to the all data
on proton-antiproton total cross sections in a whole range of
energies available up today. The global fit was made with the
function of the form
\be
\sigma^{tot}_{p\bar p}(s) = \sigma^{tot}_{asmpt}(s) \left[1 +
\frac{c}{\sqrt{s-4m^2_N}R^3_0(s)} \left(1+\frac{d_1}{\sqrt{s}} +
\frac{d_2}{s} + \frac{d_3}{s^{3/2}}\right)\right] \label{50}
\ee
where $m_N$ is proton (nucleon) mass,
\be
R^2_0(s) = \left[0.40874044 \sigma^{tot}_{asmpt}(s)(mb) -
B(s)\right](GeV^{-2}),\label{51}
\ee
\be
\sigma^{tot}_{asmpt}(s) = 42.0479 + 1.7548
\ln^2(\sqrt{s}/20.74),\label{52}
\ee
\be
B(s) = 11.92 + 0.3036 \ln^2(\sqrt{s}/20.74),\label{53}
\ee
$c, d_1, d_2, d_3$ are free parameters. Function (\ref{50})
corresponds to the structure given by Eq. (\ref{33}). 

In fact we have for the function $\chi (s)$ in the R.H.S.
of Eq. (\ref{33}) the theoretical expression in the
form\footnote{(\ref{54}) follows
from (\ref{25}) if $\delta\sigma_0(s)\sim C R_0^2(s)$ and we assume
that $R_0^2(s)\ll R_d^2$ which is justified up to extremely high
energies.}
\be
\chi (s) = \frac{C}{\kappa (s)R_0^3(s)}, \label{54}
\ee
where
\be
\kappa^4 (s) = \frac{1}{2\pi}\int_a^b dx
\sqrt{(x^2-a^2)(b^2-x^2)[(a+b)^2-x^2]},\quad a = 2m_N,\quad b =
\sqrt{2s+m_N^2}-m_N.\label{55}
\ee
It can be proved that $\kappa (s)$ has the following
asymptotics\footnote{Integral in R.H.S. of Eq. (\ref{55}) can be
expressed in terms of the Appell function.}
\[
\kappa (s)\sim \sqrt{s},\quad s\rightarrow \infty; \qquad
\kappa (s)\sim \sqrt{s-4m_N^2},\quad s\rightarrow 4m_N^2.
\]
We used at the moment the simplest function staying in the R.H.S. of
Eq. (\ref{50}) which described these two asymptotics, but at this
price we introduced the preasymptotic terms needed to describe the
region of middle (intermediate) energies.

Our fit yielded
\[
d_1 = (-12.12\pm 1.023)GeV,\quad d_2 = (89.98\pm 15.67)GeV^2,
\]
\be
d_3 = (-110.51\pm 21.60)GeV^3,\quad c = (6.655\pm
1.834)GeV^{-2}.\label{56}
\ee
The fitting curve is shown in Fig. 4.
\begin{center}
\begin{picture}(288,194)
\put(15,10){\epsfbox{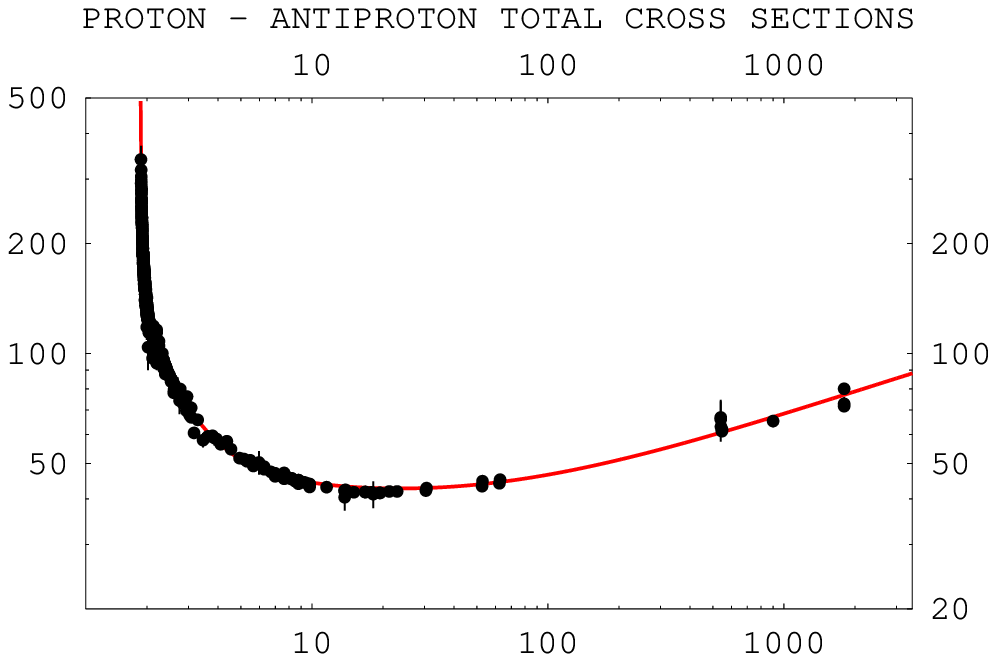}}
\put(70,75){\epsfxsize=5cm \epsfbox{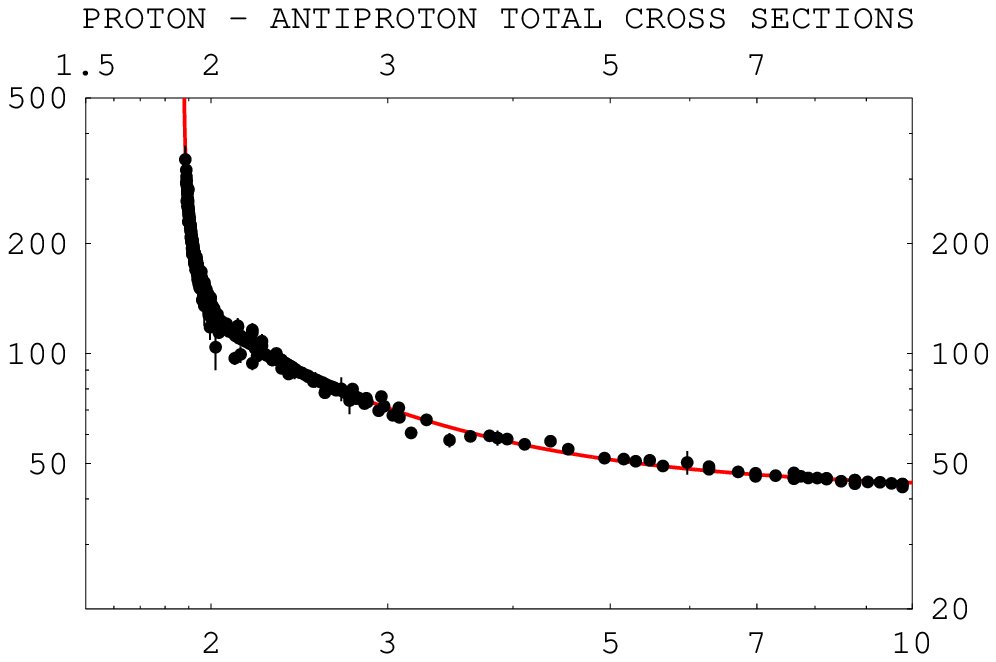}}
\put(144,0){$\sqrt{s}\, (GeV)$}
\put(0,87){\rotate{\large$\sigma_{tot} (mb)$}}
\end{picture}
\end{center}
Figure 4: The total proton-antiproton cross sections
versus $\sqrt{s}$ compared with formula (\ref{50}). Solid line
represents our fit to the data.

\bigskip
The experimental data on proton-proton total cross sections display a
more complex structure at low energies than the proton-antiproton
ones. To describe this complex structure we, of course, have to
modify formula (\ref{50}) without destroying the general structure
given by Eq. (\ref{33}). Modified formula looks like
\be
\sigma_{pp}^{tot}(s) = \sigma^{tot}_{asmpt}(s) 
\left[1 + \left(\frac{c_1}{\sqrt{s-4m^2_N}R^3_0(s)} -
\frac{c_2}{\sqrt{s-s_{thr}}R^3_0(s)}\right)\left(1 + d(s)\right) +
Resn(s)\right],\label{57}
\ee
where $\sigma^{tot}_{asmpt}(s)$ is the same as in the
proton-antiproton case (Eq. (\ref{52})) and
\be
d(s) = \sum_{k=1}^{8}\frac{d_k}{s^{k/2}},\quad Resn(s) =
\sum_{i=1}^{8}\frac{C_R^i s_R^i
{\Gamma_R^i}^2}{\sqrt{s(s-4m_N^2)}[(s-s_R^i)^2+s_R^i{\Gamma_R^i}^2]}.
\label{58}
\ee
Compared to Eq. (\ref{50}) we introduced here an additional term
$Resn(s)$ describing diproton resonances which have been extracted
from \cite{31,32}. The positions of resonances and their
widths are listed in Table I.

\bigskip
\centerline{Table I: Diproton resonances extracted from papers
\cite{31,32}.}
\begin{center}
\begin{tabular}{|c|c|c|}\hline   
$ m_R(MeV) $ & $\Gamma_R(MeV) $ & $C_R(GeV^2)$  \\ \hline     
$ 1937\pm 2 $ & $ 7\pm 2 $ & $ 0.0722\pm 0.0235 $ \\ \hline
$ 1955\pm 2 $ & $ 9\pm 4 $ & $ 0.1942 \pm 0.0292 $ \\ \hline
$ 1965\pm 2 $ & $ 6\pm 2 $ & $ 0.1344 \pm 0.0117 $ \\ \hline
$ 1980\pm 2 $ & $ 9\pm 2 $ & $ 0.3640 \pm 0.0654 $ \\ \hline
$ 2008\pm 3 $ & $ 4\pm 2 $ & $ 0.3234 \pm 0.0212 $ \\ \hline
$ 2106\pm 2 $ & $11\pm 5 $ & $-0.2958 \pm 0.0342 $ \\ \hline
$ 2238\pm 3 $ & $22\pm 8 $ & $ 0.4951 \pm 0.0559 $ \\ \hline
$ 2282\pm 4 $ & $24\pm 9 $ & $ 0.0823 \pm 0.0319 $ \\ \hline
\end{tabular}
\end{center}
The $c_1, c_2, s_{thr}, d_i, C_R^i
(i=1,...,8)$ were considered as free fit parameters. Fitted
parameters obtained by  fit are listed below (see $C_R^i$ in Table I)
\[
c_1=(192.85\pm 1.68) GeV^{-2},\quad c_2=(186.02\pm 1.67) GeV^{-2},
\]
\[
s_{thr}=(3.5283\pm 0.0052) GeV^2,
\]
\[
d_1=(-2.197\pm 1.134)10^2 GeV,\quad d_2=(4.697\pm 2.537)10^3 GeV^2,
\]
\[
d_3=(-4.825\pm 2.674)10^4 GeV^3,\quad d_4=(28.23\pm 15.99)10^4 GeV^4,
\]
\[
d_5=(-98.81\pm 57.06)10^4 GeV^5,\quad d_6=(204.5\pm 120.2)10^4 GeV^6,
\]
\be
d_7=(-230.2\pm 137.3)10^4 GeV^7,\quad d_8=(108.26\pm 65.44)10^4
GeV^8.\label{59}
\ee
The fitting curve is shown in Fig. 5. It should be pointed out
that our fit revealed the fact that the resonance with the mass
$m_R=2106\, MeV$ should be odd parity. Our fit indicates that this
resonance is strongly confirmed by the set of experimental data on
proton-proton total cross sections. That is why a further study of
diproton resonances is very desirable. 
\begin{center}
\begin{picture}(288,204)
\put(15,10){\epsfbox{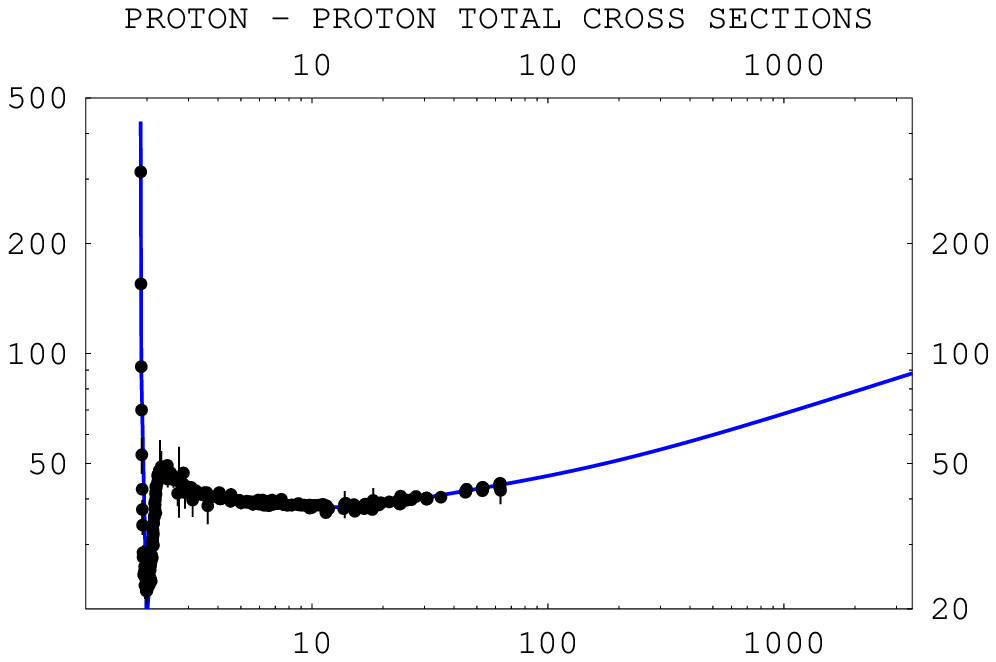}}
\put(70,75){\epsfxsize=5cm \epsfbox{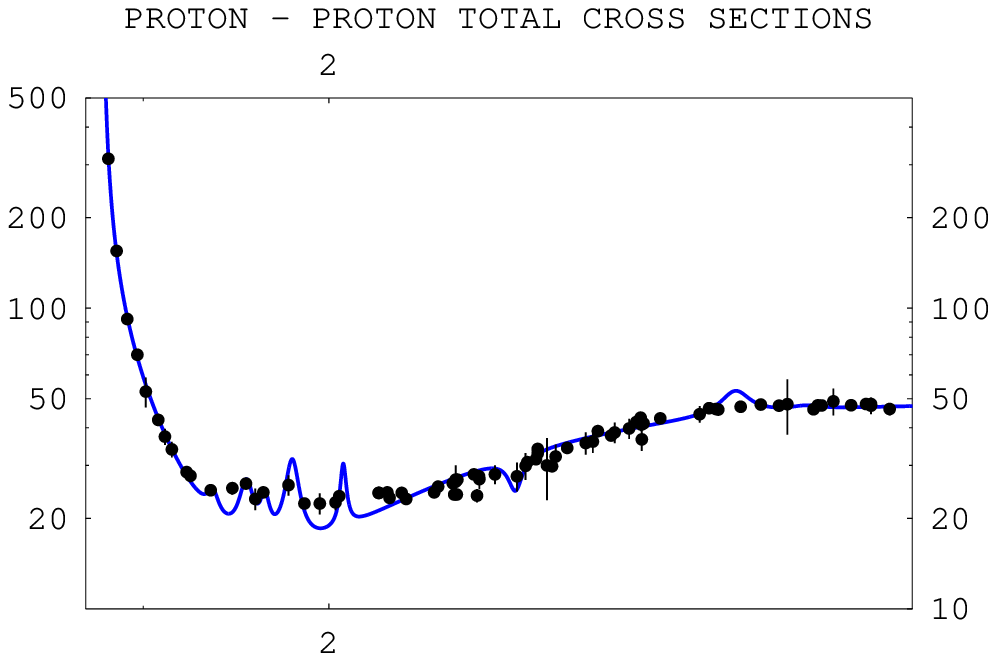}}
\put(144,0){$\sqrt{s}\, (GeV)$}
\put(0,87){\rotate{\large$\sigma_{tot} (mb)$}}
\end{picture}
\end{center}
Figure 5: The total proton-proton cross sections versus $\sqrt{s}$
compared with formula (\ref{57}). Solid line represents our fit
to the data. Statistical and systematic errors added in quadrature.

\bigskip
Figures 4, 5 display a very good correspondence of theoretical
formula (\ref{33}) to the existing experimental data on proton-proton
and proton-antiproton total cross sections. 

\section{Conclusion}

In conclusion we'd like to emphasize the
following attractive features of formula (\ref{33}). This formula
represents hadronic total cross section in a factorized form. One
factor describes high energy asymptotics of total cross section and
it has the universal energy dependence predicted by the Froissart
theorem. The other factor is responsible for the behaviour of total
cross section at low energies and this factor has also the universal
asymptotics at elastic threshold. It is a remarkable fact that the
low energy asymptotics of total cross section at elastic threshold is
dictated by high energy asymptotics of three-body (three-nucleon in
that case) forces.  This means that we undoubtedly faced very deep
physical phenomena here. The appearance of new threshold
$s_{thr}=3.5283\, GeV^2$ in the proton-proton channel, which is near
the elastic threshold, is nontrivial fact too. It's clear that the
difference of two identical terms with different thresholds in R.H.S.
of Eq. (\ref{57}) is a tail of the crossing symmetry which is not
actually taken into account in our consideration. What physical
entity does this new threshold corresponds to? This interesting
question is still open. 

Anyway we have established that simple theoretical formula (\ref{33})
described the global structure of $pp$ and $p\bar p$ total cross
sections in the whole range of energies available up today. We have
shown that this formula follows from the generalized
asymptotic theorems a l\`a Froissart. It is very nice that the
understanding of ``soft" physics based on general principles of QFT,
such as analyticity and unitarity, together with dynamic apparatus of
single-time formalism in QFT, corresponds so fine to the
experimentally observable picture. 

\acknowledgments
I am indebted to V.V. Ezhela for the access to the computer readable
files on total proton-proton and proton-antiproton cross sections in
IHEP COMPAS database. The friendly encouragement and many pieces of
good advice on computer usage from A.V. Razumov are gratefully
acknowledged.

\end{document}